\documentclass[lettersize,journal]{IEEEtran}
\usepackage{amsmath,amsfonts}
\usepackage{algorithmic}
\usepackage{algorithm}
\usepackage{array}
\usepackage[caption=false,font=footnotesize]{subfig}
\usepackage{textcomp}
\usepackage{stfloats}
\usepackage{url}
\usepackage{verbatim}
\usepackage{graphicx}
\usepackage{cite}

\usepackage{wasysym} 
\usepackage{multirow} 
\usepackage{booktabs}
\usepackage{xurl}
\usepackage{hyperref}

\usepackage[most]{tcolorbox}

\newtcolorbox{insightbox}[1]{
    enhanced,
    colback=gray!6,
    colframe=gray!45,
    arc=0.8mm,
    boxrule=0.5pt,
    left=2mm, right=2mm, top=1mm, bottom=1mm,
    fontupper=\small,
    fonttitle=\small\bfseries,
    title={#1.},
    coltitle=black,
    attach title to upper={\quad},
    before skip=5pt,
    after skip=5pt
}

\usepackage{xcolor}

\begin{document}


\title{{Beyond GDPR: Examining Disclosure Gaps in Mobile AR Privacy Policies under U.S. State Privacy Laws}}

\author{Hong~Chen,
        Xueling~Zhang,
        Hong-Ning Dai,
        Huashan~Chen,
        Qin~Yu,
        Tiange~Xie,
        Duohe~Ma, 
        and~Feng Liu%
\thanks{Hong Chen, Huashan Chen, Qin Yu, Tiange Xie, Duohe Ma and Feng Liu are with the Institute of Information Engineering, Chinese Academy of Sciences, Beijing, China, and also with School of Cyber Security, University of Chinese Academy of Sciences, Beijing, China (e-mail: chenhong2024@iie.ac.cn; chenhuashan@iie.ac.cn; yuqin@iie.ac.cn; xietiange@iie.ac.cn; maduohe@iie.ac.cn; liufeng@iie.ac.cn).}%
\thanks{Xueling Zhang is with the Rochester Institute of Technology, New York, USA
(e-mail: xueling.zhang@rit.edu).}
\thanks{Hong-Ning Dai is with the Hong Kong Baptist University, Hong Kong, China (e-mail: henrydai@comp.hkbu.edu.hk).}%
\thanks{Corresponding author: Huashan Chen.}%
}

\markboth{IEEE Transactions on Information Forensics and Security}%
{Chen \MakeLowercase{\textit{et al.}}: Beyond GDPR: Examining Disclosure Gaps in Mobile AR Privacy Policies under U.S. State Privacy Laws}

\maketitle

\begin{abstract}
Mobile Augmented Reality (MAR) apps can collect and process highly sensitive data such as spatial maps and biometrics, yet their privacy policies remain largely understudied. Prior audits of app privacy policies have typically focused on a single legal framework, such as the GDPR. Meanwhile, 20 U.S. states have comprehensive privacy laws in effect, creating a fragmented and rapidly evolving set of privacy policy obligations. To date, no study has systematically audited privacy policies against this emerging body of state-level legislation.

In this paper, we present the first large-scale audit of MAR privacy policies under U.S. state privacy laws. We construct a dataset covering the MAR ecosystem, including 8,013 Google Play MAR app metadata records worldwide, and a U.S.-based subset with 6,620 APKs and 6,426 privacy policy files. We further derive an auditable disclosure taxonomy with 5 baseline requirements, 10 triggered requirements, and 4 logic chains, and build a validated four-stage automated pipeline that produces traceable, evidence-grounded disclosure judgments.

Our audit reveals widespread disclosure gaps: 44.62\% of audited policies exhibit severe disclosure omissions, with each missing more than eight requirements, and four privacy-policy requirements have violation rates above 90\%. These findings suggest that MAR privacy disclosures are not keeping pace with the growing complexity of U.S. state privacy regulation. We release our dataset, taxonomy, and auditing pipeline to support future research on scalable privacy compliance auditing.
\end{abstract}


\vspace{-0.3cm}

\section{Introduction}
\IEEEPARstart{A}{ugmented} Reality (AR) has become a prominent form of immersive computing, overlaying digital content onto users' perception of the physical world. Among current AR form factors, Mobile AR (MAR) is the most widely accessible and practically deployed, building on ubiquitous AR-capable smartphones and mature mobile developer ecosystems that make AR experiences broadly available. Estimates place the number of MAR users worldwide at approximately 1.07 billion by 2025~\cite{statista_mobile_ar_users}. 
Unlike conventional mobile apps, MAR apps continuously sense and interpret users' physical surroundings through sensor fusion, including camera streams, GPS data and motion tracking trajectories that can reveal high-resolution visual context and biometric-adjacent information \cite{HarborthP2021MAR}.
As a result, transparency in MAR privacy policies becomes a practical disclosure concern: users and regulators must be able to understand what information is collected, how it is used, and what rights and controls are available.

The privacy risks associated with MAR are concrete rather than hypothetical. In recent years, AR filters (e.g., Snapchat\cite{Snapchat} and TikTok\cite{Tiktok}) and virtual try-on (Charlotte Tilbury\cite{CharlotteTilbury}, Louis Vuitton\cite{LouisVuitton}) have faced class-action lawsuits alleging inadequate notice, consent, or disclosure for biometric-data practices under laws such as the Illinois Biometric Information Privacy Act (BIPA). Prior work has also shown that MAR apps can engage in privacy-invasive behaviors and collect sensitive user information in ways that are difficult for users to detect~\cite{Lehman2022riksofMAR}. In this setting, privacy-policy disclosure is not merely a documentation formality: inadequate disclosure may expose developers to legal risk and limit users' ability to understand and control sensitive data practices.

At the same time, the U.S.\ legal environment governing privacy disclosures has changed substantially. In the absence of a comprehensive federal privacy law, many states have enacted their own comprehensive consumer privacy statutes. As of January~1, 2026, 20 U.S.\ states have such laws in effect~\cite{iapp_state_privacy_tracker_2026}. 
These laws impose overlapping but non-identical privacy-policy disclosure obligations, complicating compliance for MAR apps distributed across the U.S., where a single privacy policy may need to address requirements that vary by state.
This fragmented legal landscape raises an unresolved empirical question for MAR: whether apps that collect sensitive sensor data and are distributed to users across states provide privacy policies that reflect these varying disclosure obligations. 

To our knowledge, no prior large-scale study has systematically examined MAR apps from a privacy measurement perspective under U.S. state privacy laws. We address this gap through the first large-scale disclosure audit of MAR privacy policies under currently enacted U.S.\ state comprehensive privacy laws. Specifically, we aim to answer the following research questions (RQs):
\begin{itemize}
  \item \textbf{RQ1:} What are the ecosystem-scale characteristics of MAR apps distributed through Google Play?
  \item \textbf{RQ2:} What privacy-policy disclosure requirements are imposed by current U.S.\ state comprehensive privacy laws, and how do these requirements vary across states?
  \item \textbf{RQ3:} To what extent do MAR app privacy policies satisfy disclosure expectations under current U.S.\ state privacy laws, particularly in disclosing AR-specific sensitive data practices?
\end{itemize}

{A rigorous systematic audit tailored to these RQs needs to resolve three methodological challenges: i) how to identify MAR apps in app stores without dedicated AR labels, ii) how to translate fragmented state statutes into comparable disclosure requirements, and iii) how to assess disclosure completeness from heterogeneous policy text while preserving traceable evidence. To address these challenges, }we first construct a MAR dataset containing 8,013 Google Play metadata records, a U.S.-based subset with 6,620 APKs, and 6,426 privacy policies. We then derive a 22-requirement taxonomy from enacted U.S.\ state comprehensive privacy laws and organize it into an auditable model with 5 baseline requirements, 10 triggered requirements, and 7 conditional requirements in 4 logic chains. Finally, we apply a Large Language Model (LLM)-based auditing pipeline to 4,116 English MAR privacy policies. Our audit reveals substantial disclosure gaps: 44.62\% of audited policies exhibit severe disclosure omissions, each missing more than eight audited requirements. These gaps are especially pronounced for rights-related disclosures and sensitive or high-risk data practices, including biometric-data processing. In summary, this paper makes the following contributions:
\begin{itemize}
  \item We construct and release the first large-scale dataset dedicated to MAR apps, combining Google Play metadata, APKs, and privacy policies. In doing so, we introduce a robust framework for large-scale, targeted app collection that generalizes beyond MAR to studies requiring semantic filtering of app functionality.

  \item We systematically analyze currently enacted U.S.\ state comprehensive privacy laws and develop a law-grounded taxonomy that maps privacy-policy disclosure obligations to baseline, triggered, and conditional categories.

  \item We design and validate an LLM-based automated auditing pipeline for large-scale privacy-policy disclosure analysis, enabling traceable audit judgments against the law-derived taxonomy.

  \item We conduct the first large-scale empirical audit of MAR privacy policies under enacted U.S.\ state privacy laws, revealing widespread disclosure gaps across baseline, triggered, and conditional requirements.
\end{itemize}

\vspace{-0.2cm}
\section{Background and Related Work}

\begin{table*}[t]
\centering
\caption{Comparison of Existing Datasets for MAR apps.}
\vspace{-0.2cm}
\label{tab:dataset_comparison}
\begin{tabular}{@{} l c l c c c c c @{}}
\toprule
\multirow{2}{*}{\textbf{Dataset}} & \multirow{2}{*}{\textbf{Collection Year}} & \multirow{2}{*}{\textbf{Platform}} & \multirow{2}{*}{\textbf{\# Metadata}} & \multicolumn{2}{c}{\textbf{\# APKs}} & \multirow{2}{*}{\textbf{\# Policy}} & \multirow{2}{*}{\textbf{Availability}} \\
\cmidrule(lr){5-6}
& & & & \textbf{VR} & \textbf{AR} & & \\ 
\midrule
Herskovitz et al. ~\cite{Herskovitz2020iOS} & 2019 & App Store          & 0    & 0      & 105    & 0      & Closed \\ 
Yang et al. ~\cite{YangZ22MAR}       & 2022 & Google Play        & 0    & 0      & 390    & 0      & Closed \\ 
Alghamdi et al. ~\cite{Alghamdi2025xrdroid}   & 2024 & Google Play        & 408    & \multicolumn{2}{c}{408 (Mixed)} & 0 & Public \\ 
Li et al. ~\cite{li2024xrzoo}         & 2024 & Cross-Platform     & 0 & 11,949 & 579    & 0      & Pending \\ 
\midrule
\textbf{Ours}         & \textbf{2025} & \textbf{Google Play} & \textbf{8,013} & \textbf{0} & \textbf{6,620} & \textbf{6,426} & \textbf{Public} \\ 
\bottomrule
\end{tabular}
\vspace{-0.5cm}
\end{table*}

{\subsection{Existing Datasets for MAR Apps}}

MAR generally refers to AR experiences delivered through mobile devices, such as smartphones and tablets. Prior studies have largely examined VR ecosystems across platforms~\cite{Trimananda2022ovrseen,Guo24OculusVR,Zhan2024vpvet}, resulting in insufficient exploration of MAR datasets. In particular, to support in-depth privacy evaluation of MAR apps, a robust MAR dataset should cover not only app-level metadata acquired from mainstream app markets but also available APK files and their corresponding privacy policies.

Existing datasets that directly study MAR apps are limited in scale or availability. Herskovitz et al.~\cite{Herskovitz2020iOS} study 105 iOS AR apps, while Yang et al.~\cite{YangZ22MAR} collect 390 AR apps from Google Play. These studies provide early insights into MAR ecosystems, but their datasets have not been publicly released and do not include privacy policies.

Other related datasets include MAR apps but are not sufficient as large-scale MAR benchmarks. XR-Droid~\cite{Alghamdi2025xrdroid} searches Google Play using XR-related keywords and obtains 408 downloadable XR apps. However, it mixes AR and VR apps and does not provide privacy policies. XRZOO~\cite{li2024xrzoo} covers multiple platforms. For mobile app stores, XRZOO identifies 443 VR and 166 AR apps on Google Play using SDK detection with LibScout~\cite{libscout}, and further derives 201 VR and 428 AR apps on iOS through heuristic matching. While XRZOO broadens platform coverage, its mobile-store MAR subset remains relatively small, and the dataset has not been publicly released as of this writing.

In summary, current datasets fail to support reproducible, large-scale privacy evaluation for MAR scenarios. As compared in Table~\ref{tab:dataset_comparison}, no existing public dataset offers large-scale MAR app metadata, downloadable APK resources, and privacy policies in a unified manner. Targeting this limitation, we build a public Google Play-based MAR dataset covering app metadata, downloadable APKs, and privacy policies, enabling comprehensive large-scale analysis of MAR ecosystem properties and privacy compliance status.

\vspace{-0.3cm}
\subsection{Privacy Law Landscape}
With the rise of the Internet and mobile computing, privacy legislation has undergone sustained modernization worldwide. Among these developments, the General Data Protection Regulation (GDPR), enforced by the European Union in 2018, has become a major reference point for modern data protection and has broadly influenced subsequent privacy legislation. Yet existing privacy compliance research remains heavily centered on the GDPR, while the relatively small body of work that moves beyond the GDPR is still typically confined to specific jurisdictions or sectoral regimes, such as the California Consumer Privacy Act (CCPA) and the Health Insurance Portability and Accountability Act (HIPAA)~\cite{Javed2024pp_review}.

Unlike jurisdictions governed by a single comprehensive privacy regime, the United States has long lacked a federal omnibus privacy law. As a result, individual states have advanced their own legislative agendas, gradually producing a complex patchwork of state-level privacy laws. The enactment of the CCPA marks the beginning of this wave of state-level comprehensive privacy legislation. This trend has spread rapidly across the country, especially since 2023, as many states have enacted or brought into effect their own comprehensive privacy statutes. These laws share a common baseline, such as requirements to disclose categories of personal information collected, processing purposes, consumer rights, and mechanisms for exercising those rights. At the same time, they differ in their treatment of sensitive personal information, biometric data, profiling, appeal rights, controller identification, update notices, and other policy-level obligations.

This fragmentation creates a measurement challenge for privacy-policy auditing. A nationally distributed app in the United States may publish a single privacy policy, yet that policy may need to satisfy disclosure obligations that vary across states. Existing compliance studies rarely model this broader state-level landscape in a systematic way, leaving open how privacy-policy obligations differ across state laws and how those differences can be translated into auditable requirements. Our work addresses this gap by analyzing enacted U.S.\ state comprehensive privacy laws and deriving a unified requirement taxonomy for large-scale auditing.

\vspace{-0.3cm}
\subsection{Automated Analysis Tools for Privacy Policies}

Automated privacy-policy analysis has long relied on machine learning, deep learning, and natural language processing. Early systems adopt supervised classifiers trained on annotated policy corpora. For example, Polisis~\cite{Harkous2018polisis} trains hierarchical classifiers using the OPP-115 dataset to categorize policy segments and their attributes. Such approaches enable scalable policy analysis, but their dependence on manually annotated corpora can limit coverage for rare or domain-specific data. Prior work has also observed that Polisis may struggle with subtle semantic distinctions, such as distinguishing explicit denials from vague or conditional statements~\cite{Qiu2023Calpric}.

A second line of work extracts structured data practices from privacy policies. PolicyLint~\cite{Andow2019policylint} represents data practices as tuples involving actors, actions, data objects, and recipients, and uses these representations to detect internal inconsistencies. Building on this advent, PoliCheck~\cite{Andow2020policheck} uses dynamic analysis to trace data flows and verify privacy disclosure consistency. However, these approaches rely heavily on Named Entity Recognition (NER) models (e.g., spaCy, BERT) to identify ``data objects" and ``receiving entities", limiting generalizability and requiring substantial additional training to recognize domain-specific terminology (e.g., Smart Home environments~\cite{Manandhar2022smarthome}, VR~\cite{Zhan2024vpvet}). Other systems based on predefined dictionaries (PolicyChecker~\cite{Xiang2023policychecker}, PolicyPulse~\cite{Adhikari2025policypluse}, and Poligraph~\cite{Cui2023poligraph}) use semantic role labeling and dependency parsing to identify privacy-relevant statements and extract meaningful semantic frames. Such methods heavily rely on manually crafted rule bases and predefined dictionaries, which limits their flexibility and adaptability. As a result, they often struggle with complex sentence structures or cross-sentence references, and non-sentence structures, such as lists and tables, leading to information omissions.

As LLMs have demonstrated superior capabilities in semantic comprehension, recent work has begun to explore LLMs for privacy-policy understanding. Mori et al.~\cite{mori2025evaluating} show that LLMs outperform average users in understanding privacy policies and can efficiently assist companies in refining policy formulations. The work most closely related to ours is Xie et al.~\cite{Xie2025evaluating}, which uses an LLM to evaluate privacy policies under modern privacy laws. However, despite incorporating multiple laws, their framework remains largely grounded in core provisions from the CCPA and GDPR. Furthermore, their approach incorporates specific requirements from the CCPA Regulations, resulting in inconsistent granularity and potentially overly stringent disclosure expectations.

Our work extends this line of research in three respects. First, we ground the audit in a law-derived taxonomy of state comprehensive privacy laws, rather than a single framework or a small set of representative provisions. Second, we focus on MAR apps, whose use of camera, location, and biometric data introduces domain-specific privacy risks. Third, we combine LLM-based requirement annotation, extraction, and normalization to enable traceable disclosure judgments at scale.

\vspace{-0.2cm}
\section{Dataset Construction} \label{sec:dataset}

\begin{figure*}[htbp]
  \centering
  \includegraphics[width=0.95\linewidth]{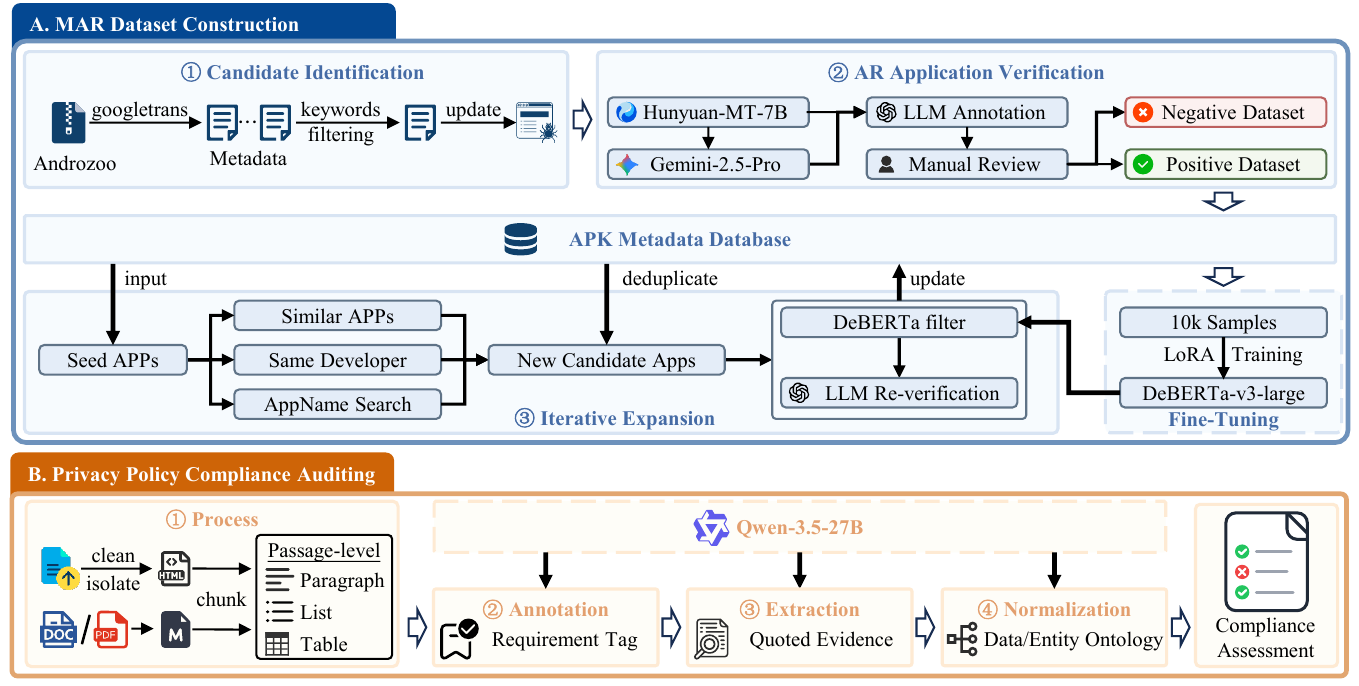}
  \caption{Overview of Pipeline.}
  \label{fig:overview_of_pipeline}
  \vspace{-0.5cm}
\end{figure*}

\subsection{Challenges and Pipeline Overview} \label{sec:dataset-overview}
Identifying MAR apps at scale presents two challenges. First, unlike VR platforms such as the Meta Quest App Store and Steam, which explicitly label apps by modality, mainstream mobile app stores (Google Play, Apple App Store) lack dedicated AR categories. MAR apps are distributed across broad store categories such as ``Photography'' or ``Games'' that do not indicate AR functionality, making direct enumeration infeasible. Second, a strategy based solely on known AR libraries is unlikely to provide sufficient coverage. Our preliminary investigation identifies over 30 distinct AR SDKs beyond well-known frameworks such as ARCore and ARKit, many of which have unstable or version-dependent signatures or limited public availability of SDK artifacts, complicating reliable detection at scale. Moreover, 24 of the SDKs we observed are absent from the 37 AR frameworks summarized by Cao et al. \cite{Cao2023MAR_SDK} as of August 2022, highlighting the challenge of maintaining a complete and up-to-date SDK inventory.

{In this study, we rely primarily on developer-provided app titles and descriptions to identify candidate MAR apps. As summarized in Fig.~\ref{fig:overview_of_pipeline}(A), we construct our MAR dataset through a three-stage pipeline: (i)~initial candidate identification from AndroZoo~\cite{Alecci2024androzoo}, (ii)~multi-stage verification combining translation models, LLMs, and human review, and (iii)~iterative expansion via breadth-first search over Google Play. We select Google Play as the focus of this study for two reasons. First, Android dominates the global mobile operating system market, accounting for 67.35\% of the market share as of April 2026, compared with 32.55\% for iOS~\cite{StatCounter2026MobileOS}. Second, {Android} provides a more feasible basis for future large-scale app-level analysis, as the closed nature of the iOS ecosystem complicates scalable app collection, program analysis, and third-party library identification~\cite{Tang2020ios}. }

\vspace{-0.3cm}
\subsection{Initial Candidate Identification} \label{sec:3.2}
We mine app metadata from AndroZoo, a repository covering over four million Google Play apps. Since app descriptions span multiple languages, we first translate all non-English titles and descriptions into English using \texttt{googletrans}~\cite{googletrans}. This preprocessing step is necessary because AR-related expressions vary substantially across languages, making direct multilingual keyword matching unreliable.

We then apply keyword matching using two patterns: (i) ``augmented reality'' in a case-insensitive manner, and (ii) ``AR'' only when it appears as a standalone token. Both patterns are applied to app titles and descriptions. This strategy rests on the assumption that developers who implement AR functionality typically mention it in their store listings. While this does not hold universally, it is effective for identifying candidate apps with AR functionality. Precision is addressed in the subsequent verification stage (Section~\ref{sec:3.3}). This keyword search yields 22,799 candidate apps after deduplication by their unique package name. Considering that AndroZoo metadata may be outdated, we re-crawl current Google Play listings for all candidates using \texttt{google-play-scraper}~\cite{googleplayscraper}. This step removes 16,891 unavailable apps, leaving \textbf{5,908} candidates for subsequent analysis.

\vspace{-0.3cm}
\subsection{AR Application Verification} \label{sec:3.3}
Notably, the 5,908 candidates obtained from Section~\ref{sec:3.2} may contain noise. For example, ``AR'' can be ambiguous, referring to entities such as Arkansas; some app descriptions mention AR only as a planned rather than an implemented feature; and some apps are educational or news-oriented apps about augmented reality rather than apps that actually provide AR functionality. To address these issues, we apply a multi-stage verification pipeline to validate all candidate apps.

\paragraph{High-quality Translation} Although \texttt{googletrans} used in Section~\ref{sec:3.2} supports keyword-based matching, it is not sufficiently reliable for semantic-level judgments. Therefore, we re-translate all descriptions using Hunyuan-MT-7B, the top-performing model in the constrained track of WMT2025~\cite{kocmi2025WMT}, which is surpassed only by large-scale closed-source models. For languages unsupported by Hunyuan-MT-7B, we use Gemini~2.5~Pro as a fallback.

\paragraph{LLM Annotation} We classify each app using GPT-4.1-mini, the latest model that supports \texttt{logprobs} in the API at the time of our study. The model takes the app name and translated description as input and is prompted with the question: ``Based on the above information, do you think this app provides augmented reality (AR) functionality?'' It outputs one of three labels: \textit{yes}, \textit{no}, or \textit{unk} (uncertain). 
Following Farr et al.~\cite{Farr2025red-ct}, we enable \texttt{logprobs}
with \texttt{top\_logprobs=3} and derive a confidence score from the first
generated token:
\[
C = \log P(t^*) -
\max_{t \in \mathcal{T} \setminus \{t^*\}} \log P(t),
\]
where $\mathcal{T}$ denotes the set of top candidate tokens, $t$ is an arbitrary token in $\mathcal{T}$, $t^*$ represents the highest-probability token, and $P(\cdot)$ denotes the probability assigned by the model. Consequently, a higher value of $C$ signifies greater confidence in the selected label.


\paragraph{Human review} We treat all \textit{unk} labels as uncertain and add the lowest-confidence \textit{yes}/\textit{no} predictions until the uncertain set reaches 10\% of all samples. Two annotators with advanced expertise in AR research and development independently label each sample as either AR or non-AR, achieving strong agreement (Cohen's $\kappa = 0.860$). Disagreements are resolved through discussion, with reference to Google Play screenshots, preview videos, and app testing when necessary. Ambiguous cases are labeled non-AR unless clear AR functionality is evident. After this verification process, \textbf{4,970} apps are confirmed as MAR apps.

\vspace{-0.3cm}
\subsection{Iterative Dataset Expansion} \label{sec:3.4}

The 4,970 verified MAR apps from Section~\ref{sec:3.3} are derived entirely from AndroZoo metadata, which does not cover all Google Play apps. To capture MAR apps absent from AndroZoo, we iteratively expand the dataset using a breadth-first search (BFS) strategy over the Google Play app graph.

\paragraph{DeBERTa fine-tuning} To reduce LLM calls during BFS expansion, we fine-tune DeBERTa-v3-large~\cite{He2023DeBERTaV3} with LoRA~\cite{Hu2022LoRA} as a cost-efficient binary classifier for AR candidate filtering. To construct high-confidence negative examples, we crawl top-free and top-paid Google Play listings and apply the verification pipeline from Section~\ref{sec:3.3}. The resulting 10,000-sample dataset is split into training, validation, and test sets at an 8:1:1 ratio. Each input pairs the app title with its description. For descriptions exceeding the context window, we use overlapping 512-token chunks with a stride of 256, prepend the title to each chunk, and compute the sample-level prediction by averaging the window-level logits. We train with weighted cross-entropy loss and select the best checkpoint by validation F1. Table~\ref{tab:deberta_results} reports the final performance. During BFS expansion, the classifier serves as a high-throughput coarse filter, forwarding only AR predictions and low-confidence rejections to the LLM for re-verification.

\begin{table}[t]
\centering
\caption{Classification performance of DeBERTa-v3-large.}
\vspace{-0.2cm}
\label{tab:deberta_results}
\begin{tabular}{lccc} 
\toprule
\textbf{Dataset} & \textbf{Precision (\%)} & \textbf{Recall (\%)} & \textbf{F1 (\%)}\\
\midrule
Validation & 97.80 & 98.45 & 98.12 \\
Test       & 99.32 & 98.00 & 98.66 \\
\bottomrule
\end{tabular}
\vspace{-0.5cm}
\end{table}

\paragraph{Expansion procedure} In each BFS round, newly discovered MAR apps are used as seeds to retrieve Google Play candidates from similar apps, other apps by the same developer, and name-based search results. After package-name deduplication, candidates are passed to a two-stage verifier:
\begin{enumerate}
    \item \textbf{DeBERTa filter.} The fine-tuned DeBERTa classifier scores each candidate and discards high-confidence non-AR apps. This threshold is empirically set to 0.96, as DeBERTa's confidence scores are right-skewed and heavily concentrated in the high-confidence interval.
    \item \textbf{LLM re-verification.} All remaining apps, including those classified as AR by DeBERTa, are forwarded to GPT-4.1-mini for verification using the same prompt described in Section~\ref{sec:3.3}. Only apps confirmed as AR by the LLM are added to the seed set for the next round.
\end{enumerate}

Expansion terminates when a full BFS round yields no new MAR apps. Overall, we crawl metadata for over 220,000 Google Play apps, expanding the dataset to \textbf{8,013} MAR apps.

\vspace{-0.3cm}
\subsection{Evaluation of MAR Identification}  \label{sec:3.5}

To evaluate the performance of our MAR identification pipeline, we construct a golden-label dataset of 300 apps via manual verification. The dataset is held out during DeBERTa fine-tuning to ensure an unbiased evaluation, and is designed to cover varied real-world cases:

\begin{itemize}
    \item \textbf{10 well-known MAR apps} whose current store descriptions do not contain AR-related keywords (e.g., Pok\'{e}mon GO, Instagram, IKEA Place).
    \item \textbf{143 positive examples}: 72 apps matching the ``AR'' keyword pattern and 71 matching the ``Augmented Reality'' pattern, all confirmed MAR apps.
    \item \textbf{147 negative examples}: (i)~48 apps containing the ambiguous token ``AR'' but not providing AR functionality, (ii)~49 apps containing ``Augmented Reality'' in their descriptions but not qualifying as MAR apps, and (iii)~50 apps containing no AR-related keywords.
\end{itemize}

We evaluate four configurations to isolate the contribution of each component. DeBERTa denotes the original model, whereas DeBERTa-FT denotes the fine-tuned DeBERTa classifier alone. LLM-only corresponds to the semantic verifier used for the initial candidate set. Cascade corresponds to the cost-aware verifier used during BFS expansion. Table~\ref{tab:pipeline_eval} reports the results. LLM-only achieves the highest F1, supporting its use for the initial verification stage. DeBERTa-FT substantially improves over the original DeBERTa baseline, showing that task-specific fine-tuning is necessary for reliable AR/non-AR classification. Cascade reduces LLM calls from 300 to 161 while maintaining high precision, making it suitable for the BFS expansion stage where LLM cost dominates.

\begin{table}[t]
\centering
\caption{AR classification evaluation on the golden label set.}
\vspace{-0.2cm}
\label{tab:pipeline_eval}
\begin{tabular}{lcccc}
\toprule
\textbf{Method} & \textbf{P (\%)} & \textbf{R (\%)} &
\textbf{F1 (\%)} & \textbf{LLM Calls} \\
\midrule
DeBERTa       & 50.51 & 98.04 & 66.67 & 0 \\
DeBERTa-FT       & 92.81 & 92.81 & 92.81 & 0 \\
Cascade       & 98.60 & 92.16 & 95.27 & 161 \\
LLM-only       & 99.34 & 98.69 & 99.02 & 300 \\
\bottomrule
\end{tabular}
\vspace{-0.5cm}
\end{table}

\vspace{-0.3cm}
\subsection{APK and Privacy Policy Collection} \label{sec:method-collection}
Starting from the identified MAR apps, we collect application binaries and privacy policy documents in two phases.

In November 2025, we use \texttt{Apkeep}~\cite{apkeep} with U.S.-region Google accounts to download APKs, successfully obtaining \textbf{6,620} of the 8,013 apps. The remaining 1,393 apps are unavailable for download: 121 are paid apps, 114 have been delisted from Google Play, 738 are region-restricted (confirmed via \texttt{SensorTower}~\cite{sensortower}, a commercial mobile app intelligence platform), and 420 are flagged as device-incompatible. 

In April 2026, we re-verify U.S. availability and confirm that 6,818 of the 8,013 apps remain downloadable in the U.S. market, indicating moderate ecosystem churn over the five-month period. We then collect privacy policies for the available apps using \texttt{Playwright}~\cite{playwright} together with \texttt{SingleFile}~\cite{singlefile} for full-page HTML capture, as SingleFile more reliably preserves dynamically rendered content, shadow DOM content, computed styles, and asynchronously loaded text than Playwright's native \texttt{page.content()} snapshot.

We manually inspect 1,035 URLs that fail automated retrieval and confirm 381 as genuinely inaccessible. The final collection contains \textbf{6,426} privacy policies: 6,354 HTML documents, 56 PDFs, 12 DOCX files, and 4 plain-text files.

\vspace{-0.3cm}
\section{Regulatory Framework Under U.S. State Privacy Legislation}

\begin{table*}[t]
\caption{Requirement coverage across five representative states. {\normalfont\CIRCLE}~= baseline, {\normalfont\LEFTcircle}~= triggered, {\normalfont\astrosun}~= conditional, {\normalfont\Circle}~= not required.}
\vspace{-0.2cm}
\label{tab:requirements}

\centering
\footnotesize
\renewcommand{\arraystretch}{1.05}

\begin{tabular*}{\textwidth}{@{\extracolsep{\fill}} c p{8.5cm} *{5}{c} @{}}
\toprule
\textbf{ID} & \textbf{Clause} & \textbf{California} & \textbf{Colorado} & \textbf{Oregon} & \textbf{New Jersey} & \textbf{Minnesota} \\
\midrule

\multicolumn{7}{c}{\textbf{Data Transparency}}\\
\addlinespace[2pt]

D1 & Scope of PI Collection
& \CIRCLE & \CIRCLE & \CIRCLE & \CIRCLE & \CIRCLE \\

D2 & Sources of PI Collected
& \LEFTcircle & \Circle  & \Circle & \Circle & \Circle \\

D3 & Scope of PI Sold/Shared/Disclosed
& \LEFTcircle & \LEFTcircle & \LEFTcircle & \LEFTcircle & \LEFTcircle \\

D4 & Third-Party Recipients of PI
& \astrosun & \astrosun & \astrosun & \astrosun & \astrosun \\

D5 & Purposes for PI Processing
& \LEFTcircle & \LEFTcircle & \LEFTcircle & \LEFTcircle & \LEFTcircle \\

D6 & PI Retention Periods/Criteria
& \LEFTcircle & \Circle & \Circle & \Circle & \LEFTcircle \\

\addlinespace[4pt]
\multicolumn{7}{c}{\textbf{Rights Notice}}\\
\addlinespace[2pt]

R1 & Right to Know / Access
& \LEFTcircle & \LEFTcircle & \LEFTcircle & \LEFTcircle & \LEFTcircle \\

R2 & Right to Correct
& \LEFTcircle & \LEFTcircle & \LEFTcircle & \LEFTcircle & \LEFTcircle \\

R3 & Right to Delete
& \LEFTcircle & \LEFTcircle & \LEFTcircle & \LEFTcircle & \LEFTcircle \\

R4 & Right to Data Portability
& \Circle & \LEFTcircle & \LEFTcircle & \LEFTcircle & \LEFTcircle \\

R5 & Right to Opt Out
& \astrosun & \astrosun & \astrosun & \astrosun & \astrosun \\

R6 & Right to Limit SPI Use
& \astrosun & \Circle & \Circle & \Circle & \Circle \\

R7 & Right to Non-Discrimination
& \LEFTcircle & \Circle & \Circle & \Circle & \Circle \\

R8 & Right to Contest Profiling
& \Circle & \Circle & \Circle & \Circle & \astrosun \\

R9 & Right to Obtain a List of Specific Third Parties
& \Circle & \Circle & \Circle & \Circle & \astrosun \\

\addlinespace[4pt]
\multicolumn{7}{c}{\textbf{Privacy Controls}}\\
\addlinespace[2pt]

PC1 & Contact Information
& \CIRCLE & \CIRCLE & \CIRCLE & \CIRCLE & \CIRCLE \\

PC2 & Appeal Mechanism
& \Circle & \Circle & \LEFTcircle & \LEFTcircle & \LEFTcircle \\

PC3 & Biometric Data Security Incident Response Protocol
& \Circle & \astrosun & \Circle & \Circle & \Circle \\

PC4 & Biometric Deletion Guidelines
& \Circle & \astrosun & \Circle & \Circle & \Circle \\

\addlinespace[4pt]
\multicolumn{7}{c}{\textbf{Policy Administration}}\\
\addlinespace[2pt]

PA1 & Last Updated Date/Effective Date 
& \Circle & \Circle & \Circle & \CIRCLE & \CIRCLE \\

PA2 & Controller Identification
& \Circle & \Circle & \CIRCLE & \Circle & \Circle \\

PA3 & Notification Process of Material Change
& \Circle & \Circle & \Circle & \CIRCLE & \CIRCLE \\

\bottomrule
\end{tabular*}
\vspace{-0.5cm}
\end{table*}

\subsection{Statutory Scope}\label{sec:framework:scope}

As of January~1, 2026, 20 U.S.\ states have comprehensive privacy laws in effect, which constitute the statutory basis for our analysis\footnote{To support transparency and reproducibility during review, we make all non-dataset research artifacts available at \url{https://anonymous.4open.science/r/artifact-C811/}. The full MAR dataset and fine-tuned DeBERTa model will be released upon acceptance for academic research purposes.}. We exclude sector-specific statutes such as HIPAA, as well as data-type-specific statutes such as BIPA. Although BIPA is an early and influential biometric privacy law, some of its disclosure-, purpose-, and retention-related requirements are now reflected in comprehensive state laws, such as Colorado's. We further omit state regulations, as detailed regulatory requirements are available only for California and Colorado, and their inclusion would result in inconsistent granularity across the analyzed legal sources.

\vspace{-0.3cm}
\subsection{Requirement Taxonomy} \label{sec:regulatory}
To answer RQ2, we translate this statutory basis into a law-grounded taxonomy of auditable privacy-policy disclosure requirements, excluding general legal duties unrelated to policy content. The extraction follows three steps:
\begin{enumerate}
    \item \textbf{Identification.} The first author reviews each statute and extracts requirements specifying what privacy policies must disclose, producing an initial pool of state-level disclosure requirements.
    \item \textbf{Consolidation.} We reconcile overlapping state requirements into a unified taxonomy through discussion and consensus, grouping functionally equivalent categories, such as Sensitive Personal Information (SPI) and Biometric Data (BD), under Personal Information (PI).
     \item \textbf{Applicability modeling.} We model requirement applicability using three categories: \emph{baseline requirements} ({\normalfont\CIRCLE}), covering minimum disclosures expected from any privacy policy; \emph{triggered requirements} ({\normalfont\LEFTcircle}), activated once a policy indicates that the app collects or processes personal information; and \emph{conditional requirements} ({\normalfont\astrosun}), applicable only under specific statutory preconditions.
\end{enumerate}

The taxonomy comprises \textbf{22 requirements} across four categories: \emph{Data Transparency} (6), \emph{Rights Notice} (9), \emph{Privacy Controls} (4), and \emph{Policy Administration} (3). Due to space constraints, Table~\ref{tab:requirements} reports five representative states that cover the observed requirement variations. 
The complete requirement mapping is provided in the Supplemental Materials. We briefly describe each category below.

\paragraph{Data Transparency (D1--D6)} These requirements cover the categories and sources of personal information (PI) collected, processing purposes, the categories of third parties involved in sharing, sale, or disclosure, and retention periods or criteria.  D1 and D3 are universal across all 20 states. D4 and D5 are required by 19 states each, whereas D2 and D6 are less common, required only by California and by California plus Minnesota, respectively. California also requires negative disclosure for certain practices, i.e., a controller must clearly state when it does not sell, share, or disclose PI. In practice, privacy policies that do not collect PI often disclose this explicitly. Therefore, we treat such negative disclosure as part of adequate disclosure under D1.

\paragraph{Rights Notice (R1--R9)}
All 20 state comprehensive privacy laws require privacy policies to inform consumers of at least some statutory rights. In our taxonomy, this category includes both the disclosure of consumer rights and general instructions for exercising them, because real-world privacy policies often present these elements together. Across states, the core rights include the rights to access, deletion, correction, data portability, and opt out of covered processing, which are required by nearly all state laws. State-specific extensions include California's rights to limit sensitive personal information use and to be free from discriminatory treatment, Minnesota's right to contest profiling decisions with legal or similarly significant effects, and Minnesota and Maryland's right to obtain a list of specific third parties receiving personal data.

\paragraph{Privacy Controls (PC1--PC4)} This category captures procedural requirements beyond merely stating rights or general exercise procedures. PC1 requires a concrete privacy contact channel, such as an email address, telephone number, mailing address, contact form, or another clearly identified communication mechanism. 
PC2 concerns an appeal mechanism for challenging a controller's decision regarding a privacy request. PC3 and PC4 are biometric-specific controls unique to Colorado, covering a biometric data security incident response protocol and biometric deletion guidelines.

\paragraph{Policy Administration (PA1--PA3)} These requirements address the privacy policy document itself rather than data-processing disclosures. PA1 concerns whether the policy provides its last updated date or effective date, and appearance in only four states: Oregon, Delaware, Minnesota, and Maryland. PA2 appears only in Oregon and concerns whether the policy identifies the data controller. PA3, imposed only by New Hampshire and Minnesota, concerns whether the policy explains how consumers will be notified of material changes.

\begin{insightbox}{Insight 1}
Although U.S. state comprehensive privacy statutes are fragmented, their policy-disclosure obligations can be consolidated into a shared auditable taxonomy while preserving state-specific differences in requirement scope, applicability, and granularity. This variation is reflected in the uneven prevalence of individual requirements: some are broadly shared across states, whereas others appear only in one or a few states.
\end{insightbox}

{We further organize conditional requirements into four logic chains. Each chain is activated when a policy discloses the corresponding prerequisite data practice or processing context:}

\begin{itemize}
    \item \textbf{Chain~1 (Third-Party Disclosure and Opt-Out):} $\text{D3} \rightarrow (\text{D4} \cup \text{R9})\cap \text{R5}$. If the controller sells, shares, or discloses personal information (PI) to third parties, the privacy policy must explicitly identify the categories of recipients or inform consumers of their right to request a list of specific third parties. Furthermore, the policy must acknowledge the consumers' right to opt out and provide an accessible mechanism.
    \item \textbf{Chain~2 (Limit SPI Use):} ($\text{SPI} \in \text{D1}) \rightarrow \text{R6}$. If the controller collects sensitive personal information (SPI) as disclosed under D1, then the policy must describe the consumer's right to limit the use of such data (R6).
    \item \textbf{Chain~3 (Biometric Data Protection):} ($\text{BD} \in \text{D1}) \rightarrow (\text{PC3} \cap \text{PC4})$. If the controller collects biometric data (BD) as disclosed under D1, then the policy must describe the security breach notification protocol (PC3) and deletion guidelines (PC4) for biometric data.
    \item \textbf{Chain~4 (Contest Profiling):} $(\text{Consumer Profile} \in \text{D1}) \rightarrow \text{R8}$. If the controller engages in consumer profiling as disclosed under D1, then the policy must describe the consumer's right to contest profiling decisions (R8).
\end{itemize}

\begin{insightbox}{Insight 2}
    U.S. state privacy legislation creates a disclosure dependency structure rather than a uniform checklist. Therefore, auditing MAR privacy policies requires modeling legal preconditions explicitly: a policy may appear complete at the requirement level yet become deficient once its own disclosures trigger downstream requirements.
\end{insightbox}

\vspace{-0.3cm}
\section{Disclosure Auditing Methodology}

We design a four-stage automated auditing pipeline for privacy-policy disclosure analysis. Fig.~\ref{fig:overview_of_pipeline}(B) summarizes the downstream auditing component, which maps raw privacy policies to structured per-app disclosure assessments based on the requirement taxonomy in Section~\ref{sec:regulatory}.

\vspace{-0.3cm}
\subsection{Stage~1: Policy Processing}
We convert all collected documents into a uniform passage-level representation through three steps: 

\emph{Clean.}
For HTML documents, we remove high-volume embedded payloads, primarily Base64-encoded \texttt{data:}~URIs for images and fonts, which can considerably inflate document size without contributing meaningful textual content. We also remove boilerplate noise while preserving the DOM structure.

\emph{Isolate.}
We apply \texttt{readability-lxml}~\cite{readabilitylxml} to isolate the main content body from the cleaned HTML, producing a structurally simplified document. We deliberately refrain from applying existing HTML-to-Markdown tools such as \texttt{html2text}~\cite{html2text}, \texttt{trafilatura}~\cite{trafilatura}, or \texttt{MarkItDown}~\cite{markitdown} at this step, because these tools often collapse large sections of text into a single paragraph, thereby losing the semantic structure (e.g., headings, lists, and table boundaries) required for downstream chunking. Since \texttt{readability-lxml} may strip \texttt{white-space} related CSS declarations, such as \texttt{pre}, \texttt{pre-line}, and \texttt{pre-wrap} during main-content extraction, we identify and restore these styles to preserve explicit line-break semantics in preformatted text blocks. The extracted content is then rewrapped into a standardized HTML template. This step is not applied to PDF, DOCX, or TXT files.

\emph{Chunk.}
This step segments isolated content into passage-level records for downstream analysis. Our segmentation strategy is document-type specific. For HTML documents, we recursively traverse the DOM tree with tag-specific handling. Container tags such as \texttt{div}, \texttt{section}, \texttt{article}, \texttt{details}, \texttt{ul}, and \texttt{ol} are recursively descended rather than emitted directly, since privacy policies often place large amounts of text inside a single container element. Paragraph tags \texttt{<p>} are emitted as paragraph passages, while inline \texttt{<br>} elements are treated as passage boundaries. List items \texttt{<li>} are emitted as list passages.

Tables are processed row by row. Rows under 512 tokens are emitted directly. For rows exceeding this threshold, cells longer than 512 tokens are recursively descended, as rows above this empirical threshold usually indicate non-standard table usage, where substantive prose is embedded in table cells. Nodes styled with \texttt{white-space: pre}, \texttt{pre-line}, or \texttt{pre-wrap} are split at explicit line breaks and converted into multiple \texttt{<p>} elements before further processing. Finally, for text blocks that encode lists inline instead of using explicit HTML list tags, we apply a heuristic restructuring rule: if a block contains at least two items with a consistent bullet style (symbolic, numeric, or alphabetic), we convert it into a leading explanatory paragraph followed by a \texttt{<ul>} list.

We use \texttt{MarkItDown} to convert PDF and DOCX files to Markdown, and segment them by blank lines. TXT files follow the same blank-line segmentation.

Finally, we apply language detection to all passage records and retain only English-language policies, yielding \textbf{4,116} privacy policies for downstream analysis. The maximum-length record is an outlier caused by erroneous CSS styling in the original developer-authored document and is left uncorrected.

\vspace{-0.3cm}
\subsection{Stage~2: Annotation}\label{sec:method-annotation}
The annotation stage assigns requirement labels to each passage, directing downstream extraction to the relevant text segments. We employ a two-level labeling scheme, as a single passage may address multiple requirements simultaneously. \emph{Coarse annotation} assigns each passage to one or more of the four parent categories from our regulatory taxonomy (Data Transparency, Rights Notice, Privacy Controls, Policy Administration). \emph{Fine-grained annotation} further assigns specific requirement labels within each parent category.



We use Qwen~3.5-27B~\cite{qwen35} as the annotation model and run it on a single NVIDIA RTX~Pro~6000 GPU with FP8 quantization under the officially recommended inference configuration. The official documentation reports near-identical performance between FP8-quantized and full-precision inference for this model family, while benchmark results~\cite{artificialanalysis_leaderboard} place Qwen~3.5-27B in a range comparable to GPT-5~mini. This single-card configuration supports annotation of passages from 4,116~privacy policies without requiring a multi-node cluster.

The annotation prompt includes the full requirement definitions from Section~\ref{sec:regulatory} and a one-shot example. We use the model in non-reasoning mode since the task is primarily requirement classification, not complex multi-step reasoning.

\vspace{-0.3cm}
\subsection{Stage~3: Extraction}\label{sec:method-extraction}

The extraction stage is applied only to annotated passages where span-level evidence is required. Specifically, it uses LLM-based extraction for Data Transparency requirements (D1--D6), whose assessment depends on identifying specific data items, entities, or purposes in the policy text, and regular expressions for Last Updated Date / Effective Date (PA1). Other requirements, such as Rights Notice, are assessed directly from annotation labels, where the presence or absence of a labeled passage determines whether the disclosure is present.

\begin{figure*}[t]
\centering
\subfloat[]{\includegraphics[width=0.32\textwidth]{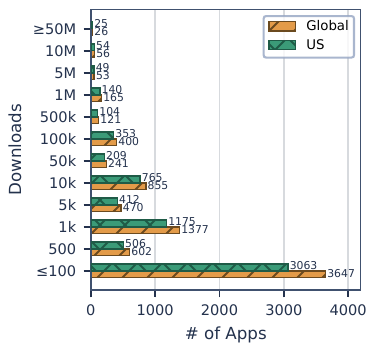}%
\label{fig:eco-downloads}}
\hfil
\subfloat[]{\includegraphics[width=0.32\textwidth]{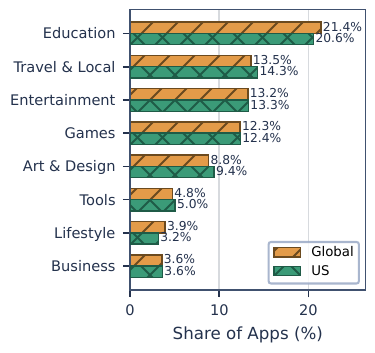}%
\label{fig:eco-category}}
\hfil
\subfloat[]{\includegraphics[width=0.32\textwidth]{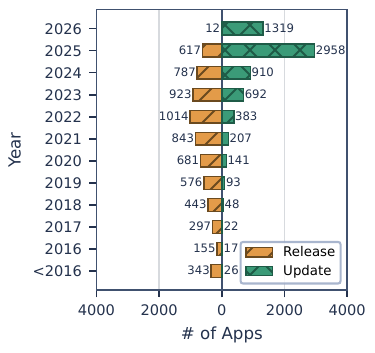}%
\label{fig:eco-years}}
\caption{Ecosystem characteristics of identified MAR apps. (a) Downloads. (b) Top-8 Categories. (c) Release vs Update Year in U.S.}
\label{fig:ecosystem-overview}
\vspace{-0.5cm}
\end{figure*}

For each passage--requirement pair in the targeted subset, the extraction model receives the annotated passage together with the requirement definition and returns exact quotes from the original policy text. To maintain extraction fidelity, we require the model to return direct textual quotations instead of paraphrased or summarized content, ensuring that the original wording remains intact for legal interpretation.
This design ensures full traceability, as every audit judgment can be traced back to a specific text span in the source policy document. We evaluate this fidelity separately in Section~\ref{sec:audit-eval}.

\vspace{-0.3cm}
\subsection{Stage~4: Normalization}\label{sec:method-normalization}

Privacy policies employ heterogeneous terminology to describe equivalent concepts. For example, ``precise location'', ``GPS data'', ``geolocation information'', and ``device location'' may all refer to the same statutory category.
To enable cross-app comparison and aggregate disclosure statistics, we normalize the extracted text spans to a law-grounded ontology derived from the definitions in all 20~state comprehensive privacy laws. The normalized outputs map heterogeneous policy expressions to canonical data-type, entity, and purpose categories, enabling aggregate disclosure statistics and conditional-chain triggers.

\vspace{-0.3cm}
\section{Results and Findings}

\subsection{Ecosystem Characteristics of MAR} \label{seg:6.1}

To answer RQ1, we characterize MAR apps distributed through Google Play along several ecosystem-scale dimensions. Across these dimensions, the U.S.-available MAR apps closely mirror the global MAR corpus, suggesting that the U.S. subset used for disclosure auditing is broadly representative of the identified MAR ecosystem.

Fig.~\ref{fig:eco-downloads} shows a highly skewed distribution of app popularity. Most MAR apps fall into low-download buckets, while only a small number reach million-level downloads. This pattern is consistent with prior studies showing that Google Play app adoption is strongly concentrated around hit applications~\cite{zhong2013googleplay}. It also motivates ecosystem-scale auditing beyond highly visible apps, since MAR includes a large population of lower-download apps. 

Fig.~\ref{fig:eco-category} presents the eight most common categories of MAR apps, collectively accounting for approximately 80\% of such apps.
\textit{Education} is the largest category, followed by \textit{Travel \& Local}, \textit{Entertainment}, and \textit{Games}. This category distribution differs from those reported in prior MAR app studies~\cite{Herskovitz2020iOS,YangZ22MAR}. 
In contrast, categories most directly associated with AR filters, namely \textit{Photography}, \textit{Social}, \textit{Beauty}, and \textit{Video Players \& Editors}, represent only a small fraction of the ecosystem. Collectively, these categories account for only 2.94\% of the global sample and 3.03\% of the U.S. sample.

\begin{insightbox}{Insight 3}
    While face-related biometric processing has drawn substantial legal and public attention, MAR privacy risks are not confined to face-centric AR applications; the dominant source of risk may instead arise from the unexpected leakage of visual information about users or bystanders in privacy-sensitive settings such as homes and outdoor spaces.
\end{insightbox}

Fig.~\ref{fig:eco-years} presents the annual release and last-update distributions of U.S.\ MAR apps. The release timeline shows that the U.S.\ MAR ecosystem accelerates after 2016 and reaches its highest annual release volume in 2022. Last-update timestamps reveal a mixed maintenance picture: as of our April 2026 collection, 76.10\% (5,187/6,816) U.S.\ MAR apps have been updated in 2024--2026, but 23.90\% (1,629/6,816) have not been updated since 2023 or earlier. The release statistics exclude 126 apps with missing release dates, while the update statistics exclude one app with a missing update timestamp.

\vspace{-0.3cm}
\subsection{Disclosure Gaps in MAR Privacy Policies} \label{seg:6.2}

\begin{figure*}[t]
\centering
\subfloat[]{\includegraphics[height=0.22\textwidth]{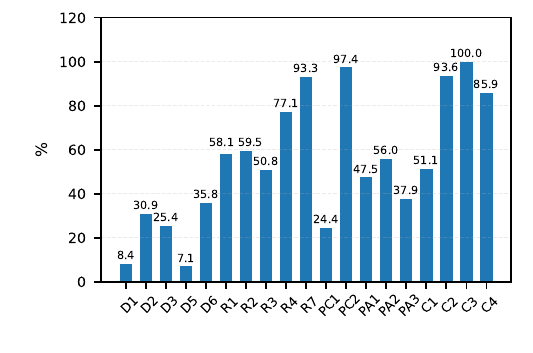}%
\label{fig:compliance-completeness}}
\hfil
\subfloat[]{\includegraphics[height=0.22\textwidth]{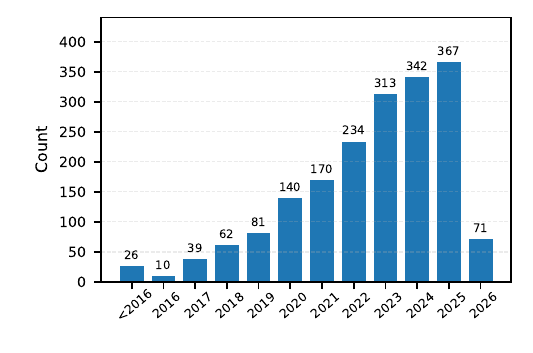}%
\label{fig:compliance-policy-dates}}
\hfil
\subfloat[]{\includegraphics[height=0.22\textwidth]{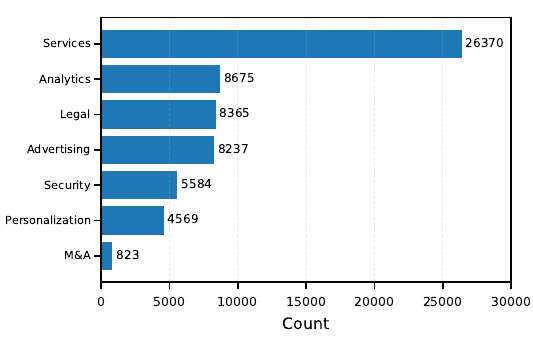}%
\label{fig:compliance-purposes}}
\caption{MAR privacy-policy audit results. (a) Violation counts. (b) Policy effective/update years. (c) Disclosed processing purposes.}
\label{fig:compliance-overview}
\vspace{-0.5cm}
\end{figure*}

To answer RQ3, we audit 4,116 eligible privacy policy files and exclude 261 unlabeled files. Manual review shows that most are non-policy pages, such as advertisements and introduction pages, while a few require multi-step navigation or span multiple policy documents. Among the remaining analyzable policies, Fig.~\ref{fig:compliance-completeness} shows that completeness violations are not concentrated in a small tail of unusually poor policies, but are widely distributed across the audited corpus. Notably, 44.62\% fall into the strong-violation range of more than eight audited requirements.

\begin{insightbox}{Insight 4}
    Disclosure incompleteness in MAR privacy policies is a recurring corpus-wide pattern. The four most frequently violated requirements all exhibit violation rates above 90\%, suggesting that these disclosure obligations have not yet been sufficiently recognized or systematically addressed within the MAR developer community, leaving many privacy notices structurally incomplete relative to current U.S. state privacy-law expectations.
\end{insightbox}

\subsubsection{Baseline requirement disclosure} We first evaluate compliance with the five baseline requirements across all 3,855 analyzable privacy policies and observe substantial omissions in these foundational disclosures. The violation rate for Scope of PI Collection is relatively low at 8.4\% ($n$ = 323), suggesting that most policies at least address whether categories of personal information are collected, as reported in Table~\ref{tab:data_type_collection_only}. However, omissions are much more common for administrative and accountability disclosures. More than half of the analyzed policies (56.0\%, $n$ = 2,157) fail to explicitly identify the data controller, and nearly half (47.5\%, $n$ = 1,830) lack a Last Updated Date or Effective Date. In addition, 37.9\% ($n$ = 1,461) omit the Notification Process of Material Change, and 24.4\% ($n$ = 941) fail to provide functional Contact Information. 

\vspace{-0.2cm}
\begin{table}[h]
\centering
\caption{Data type distributions across collection scopes.}
\vspace{-0.2cm}
\label{tab:data_type_collection_only}
\begin{tabular}{lrr}
\toprule
\multirow{2}{*}{Data Type} & \multicolumn{2}{c}{\textbf{Collection}} \\
\cmidrule(lr){2-3}
& 1st & 3rd \\
\midrule
Personal Identifier & 7229 & 970 \\
Internet/Electronic Network Activity & 4803 & 1199 \\
Data (Unspecified) & 3405 & 499 \\
Sensitive Data & 2252 & 255 \\
Coarse/Approximate Location & 927 & 153 \\
Sensory Data & 823 & 110 \\
Commercial Information & 647 & 87 \\
Consumer Profile & 500 & 101 \\
Biometric Data & 514 & 87 \\
Professional/Employment-Related Information & 444 & 18 \\
\midrule
\textbf{Total} & \textbf{21544} & \textbf{3479} \\
\bottomrule
\end{tabular}
\vspace{-0.25cm}
\end{table}

To further investigate the temporal validity of these policies, we use regular expressions to extract the most recent date from each policy, yielding 1,855 valid dates for statistical analysis; 3 additional documents are excluded due to malformed date strings or anomalous values. Fig.~\ref{fig:compliance-policy-dates} shows the distribution of these extracted policy dates. However, the mere inclusion of a Last Updated Date or Effective Date does not necessarily indicate that a policy satisfies update-frequency expectations.
California and Florida require privacy policies to be updated at least once every 12 months. Only 438 of the 1,855 policies with valid extracted dates meet this stricter recency criterion.

\begin{insightbox}{Insight 5}
    MAR privacy policies often function as static artifacts rather than living documents. When policy maintenance fails to keep pace with app updates, shifting data practices, and evolving legal expectations, disclosure gaps emerge and legacy disclosures quickly become outdated.
    
\end{insightbox}

\subsubsection{Triggered requirement disclosure} We next exclude 47 privacy policies that contain only baseline labels and explicitly state that no personal data are collected.

Policies show comparatively higher levels of completeness under Data Transparency. Only 7.1\% ($n$ = 269) fail to disclose the purposes of PI processing, with service provision appearing as the most common disclosed purpose (Fig.~\ref{fig:compliance-purposes}). 25.4\% ($n$ = 966) do not explicitly state whether they sell, share, or disclose PI. The gaps become more pronounced for provenance and lifecycle information: 30.9\% ($n$ = 1,177) fail to identify the sources of PI collected, and 35.8\% ($n$ = 1,362) do not provide specific PI retention periods or retention criteria.

The largest gaps appear in rights-related disclosures. More than half of the policies fail to mention core consumer rights, including the Right to Delete (50.8\%, $n$ = 1,935), the Right to Know/Access (58.1\%, $n$ = 2,213), and the Right to Correct (59.5\%, $n$ = 2,264). Policies perform worse on newer or more specialized disclosures: 77.1\% ($n$ = 2,937) fail to mention the Right to Data Portability, 93.3\% ($n$ = 3,551) omit the Right to Non-Discrimination, and 97.4\% ($n$ = 3,709) fail to describe an Appeal Mechanism.


\subsubsection{Conditional requirement disclosure} We assess the four conditional logic chains defined in Section~\ref{sec:regulatory} and find that violations are widespread.

\textbf{Third-Party Disclosure and Opt-Out (Chain~1).} Among the 1,508 policies that disclose selling, sharing, or disclosing personal information, most identify third-party recipients, as shown in Table~\ref{tab:source-vs-recipients}. Only 4.84\% ($n$ = 73) fail to do so. However, 51.13\% ($n$ = 771) fail to provide the corresponding Right to Opt Out. This gap suggests that many policies disclose third-party data transfers while omitting the user control required to limit or prevent such transfers.

\vspace{-0.2cm}
\begin{table}[h]
  \centering
  \caption{Source vs. recipient counts by category.}
  \vspace{-0.2cm}
  \label{tab:source-vs-recipients}
  \begin{tabular}{lrr}
    \toprule
    Category & Source & Recipients \\
    \midrule
    First-Party & 1903 & 200 \\
    Consumer & 5145 & 409 \\
    Affiliates & 115 & 1500 \\
    Third-Party (Unspecified) & 2228 & 7837 \\
    Advertising Networks & 139 & 1629 \\
    Analytics Providers & 273 & 1852 \\
    Authentication Providers & 186 & 126 \\
    Content Providers & 41 & 300 \\
    Data Brokers & 83 & 87 \\
    Email Service Providers & 101 & 208 \\
    Government Entities & 50 & 1196 \\
    Internet Service Providers & 50 & 77 \\
    Operating Systems/Platforms & 699 & 128 \\
    Payment Processors & 145 & 511 \\
    SDK Providers & 39 & 121 \\
    Social Networks & 727 & 637 \\
    \bottomrule
  \end{tabular}
\end{table}

\textbf{Biometric Data Protection (Chain~3).} The largest conditional gap appears in biometric data processing. Among the 601 policies that disclose biometric-data collection, 99.8\% ($n$ = 600) lack a Biometric Data Security Incident Response Protocol, and 97.2\% ($n$ = 584) fail to provide Biometric Deletion Guidelines. This indicates a disconnect between disclosed collection of high-risk biometric data and the corresponding security and lifecycle-management disclosures.

\textbf{Sensitive and Automated Processing (Chain~2 and~4).} Similar gaps appear for other advanced data-processing practices. Among the 1,918 policies disclosing sensitive personal information (SPI), 93.6\% ($n$ = 1,796) fail to notify users of their Right to Limit SPI Use. Among the 903 policies disclosing consumer profiling, 85.9\% ($n$ = 776) do not provide users with the Right to Contest Profiling.

\begin{insightbox}{Insight 6}
    MAR privacy policies more consistently address Data Transparency than Rights Notice, Privacy Controls, or Policy Administration. This imbalance suggests that these policies function primarily as notices of data practices, rather than as user-facing documents that support understanding, rights awareness, and meaningful control over data processing.
    
\end{insightbox}

\vspace{-0.3cm}
\subsection{Auditing Pipeline Evaluation} \label{sec:audit-eval}

\subsubsection{Extraction fidelity validation}
Since hallucinations in LLMs have long been a major concern in the academic community~\cite{Schimanski24llm_specialists, ZhangZ25llm_hallucinations_in_code}, we validate whether Stage~3 outputs are faithful to the source policy rather than being hallucinated or paraphrased. For each extracted text span, we perform a sliding-window search over the source sentences to identify the top-1 candidate window using \texttt{RapidFuzz}~\cite{rapidfuzz}. We then apply a two-tier scoring procedure:

\begin{enumerate}
  \item \emph{Literal containment shortcut.} Both the extracted span and the candidate window are normalized. If the normalized extracted span is a substring of the normalized candidate, the similarity score is set to~1.0.
  \item \emph{Embedding similarity fallback.} When literal containment does not hold, we encode both spans using Qwen3-Embedding-0.6B~\cite{Zhang2025qwen3embedding}. The final score is the cosine similarity between the two vectors.
\end{enumerate}

Table~\ref{tab:extraction-fidelity} shows extraction fidelity at two similarity thresholds (0.85, 0.95), from lenient to strict matching. Even at the strictest threshold of 0.95, 98.32\% of extracted spans match the source text, indicating that the extraction stage rarely produces hallucinated or heavily paraphrased evidence.

\vspace{-0.2cm}
\begin{table}[h]
    \centering
    \caption{Extraction fidelity at different similarity thresholds.}
    \vspace{-0.2cm}
    \label{tab:extraction-fidelity}
    \begin{tabular}{lrrr}
        \toprule
        \textbf{Label} & \textbf{Total} & \textbf{$\ge 0.85$} & \textbf{$\ge 0.95$} \\
        \midrule
        Scope of PI Collection & 134,646 & 98.71\% & 97.93\% \\
        Sources of PI Collected & 32,945 & 99.27\% & 98.56\% \\
        Scope of PI Sold/Shared/Disclosed & 36,301 & 99.36\% & 98.63\% \\
        Third-Party Recipients of PI & 62,767 & 98.99\% & 98.15\% \\
        Purposes for PI Processing & 142,278 & 99.52\% & 98.74\% \\
        PI Retention Periods/Criteria & 19,112 & 99.24\% & 97.49\% \\
        \midrule
        \textbf{TOTAL} & \textbf{429,049} & \textbf{99.14\%} & \textbf{98.32\%} \\
        \bottomrule
    \end{tabular}%
\end{table}

\subsubsection{Audit judgment evaluation}
To evaluate the audit judgments produced by the auditing pipeline, we construct a gold-standard dataset through manual annotation. We use stratified sampling based on the number of violations to cover policies with different disclosure-completeness levels, randomly selecting 100 privacy policies from our corpus.

The same two annotators independently label these policies using the pipeline's labeling scheme, producing 5,749 annotation records. Inter-rater reliability indicates substantial agreement for the binary task of identifying privacy-relevant provisions, with an agreement rate of 90.42\% and Cohen's $\kappa$ of 0.806. For the multi-label requirement-labeling task, the annotators achieve an Exact Match rate of 73.40\% (4,220/5,749), with an average Jaccard of 80.88\%, Micro-F1 of 82.59\%, and Macro-F1 of 79.13\%. Discrepancies in 1,529 annotations are adjudicated through discussion among all the annotators.

\begin{table}[t]
\centering
\caption{Performance of the automated auditing pipeline}
\vspace{-0.2cm}
\label{tab:performance_results}
\begin{tabular}{lccc}
\toprule
\textbf{Binary} & \textbf{Precision (\%)} & \textbf{Recall (\%)} & \textbf{F1-Score (\%)} \\
\midrule
Passage  & 92.78 & 82.70 & 87.45 \\

\midrule\midrule 
\textbf{Multi-label} & \textbf{Jaccard (\%)} & \textbf{Micro-F1 (\%)} & \textbf{Macro-F1 (\%)} \\
\midrule
Passage & 73.88 & 74.39 & 66.78 \\
Document & 73.56 & 87.69 & 83.98 \\
\bottomrule
\end{tabular}
\vspace{-0.5cm}
\end{table}

\vspace{-0.3cm}
\section{Discussion}

Table~\ref{tab:performance_results} presents the performance of our automated framework on the human-adjudicated ground truth. In the binary filtering task, the framework achieves 92.78\% precision, 82.70\% recall, and 87.45\% F1, indicating that it can reliably identify privacy-relevant passages while filtering out irrelevant policy text.

For the fine-grained multi-label task, we report both document-level and passage-level performance. The document-level results are the primary validity indicators of audit reliability, because our main findings are computed at the policy level: a requirement is considered disclosed if the policy contains at least one passage that satisfies the corresponding requirement. As shown in Table~\ref{tab:performance_results}, the framework achieves strong document-level performance, indicating that it can reliably recover the set of requirements disclosed by each policy.

Passage-level scores mainly reflect evidence localization granularity, and this labeling is substantially stricter because it requires assigning requirement labels to the exact text segments where the disclosure appears. The difficulty of this task is also reflected in the human annotations: the annotators achieve an Exact Match rate of 73.40\%, meaning that more than one quarter of passage-level label sets differ between annotators before adjudication. This disagreement highlights the inherent ambiguity of boundary setting and label assignment in passage-level multi-label privacy-policy analysis. Against this backdrop, the framework's passage-level results indicate that it captures much of the expert labeling signal, while remaining subject to the same boundary and interpretation challenges observed in human annotation.

We manually analyze passages with label disagreements to better understand the model's error modes. Of the 1,999 mismatches, under-labeling dominates (1,183; 59.17\%), followed by over-labeling (473; 23.66\%) and label misalignment (343; 17.16\%). A closer examination shows that false negatives largely arise when a passage does not explicitly specify the underlying data type (627) or entity (332). When a policy refers only abstractly to ``information'', ``data'' or ``third parties'',
the model tends to be more conservative than human annotators in mapping the passage to fine-grained requirement labels.

Over-labeling and label misalignment point to two additional sources of divergence. First, some privacy policies lack standardized structural markup, such as correct HTML \texttt{<li>} tags for lists or explicit heading indicators. As a result, section headers and list items can be split into context-poor passage fragments, leading experts and the model to assign labels at different passage boundaries. Second, some policies are written primarily in GDPR-oriented terminology and structure, while our taxonomy is derived from U.S.\ state privacy laws. This legal-regime mismatch can lead to different interpretations when a policy discloses a concept that is related to, but not directly aligned with, a specific U.S.\ state-law requirement.

\vspace{-0.3cm}
\subsection{Implications and Recommendations}

Our findings suggest that MAR privacy policies are often maintained as static documentation artifacts rather than as living documents that evolve with app functionality, data practices, and legal obligations due to developers' lack of attention to privacy obligations. This interpretation is supported by the observed staleness of many privacy policies relative to app update activity (Section~\ref{seg:6.2}), and by a secondary analysis of template-based policy generation. Using keyword fingerprinting inspired by Pan et al.~\cite{Pan2024APPGs}, we find that 11.92\% (766/6,426) of policies in our full dataset are highly likely to have been generated by Automated Privacy Policy Generators (APPGs), with the proportion rising to 16.38\% (674/4,116) among English policies. These policies perform only marginally better than manually drafted policies and still exhibit substantial deficiencies: 56.38\% contain between 5 and 8 violations, and 23.74\% contain more than 8 violations. These results suggest that generic policy templates may help developers cover basic disclosures, but they are often insufficient for satisfying the more specific and conditional obligations imposed by modern state privacy laws.

To bridge these gaps, we propose the following actionable recommendations for key stakeholders.

\textbf{For regulators and policymakers.}
Our findings highlight the need to refine privacy frameworks for immersive and sensor-rich apps like MAR. In particular, regulators should carefully consider whether existing definitions of Sensitive Personal Information adequately capture the privacy risks introduced by MAR. Mainstream MAR apps are not concentrated in categories such as Photography, Social Media, or Beauty, which are more commonly associated with biometric or facial-related processing. Instead, they are found in fields such as Education, Travel \& Local, Entertainment, and Games (Section~\ref{seg:6.1}). These apps may continuously process environmental and spatial data, such as depth data, high-resolution visual streams, spatial maps, and surrounding-object or bystander information. Such data may not always be explicitly categorized as Sensitive data, but it can still reveal sensitive information about users, bystanders, and private physical environments. Regulators should therefore consider extending or clarifying sensitive-data protections to cover continuous environmental and spatial data generated by MAR systems, beyond narrowly defined biometric identifiers.

\textbf{For app markets and platforms.}
Our results also point to a platform-level enforcement gap. Google Play's Developer Policy requires apps to disclose how they access, collect, use, and share user data, including developer or controller information, privacy contact mechanisms, categories of personal and sensitive data processed, third-party sharing practices, secure data handling procedures, and data retention and deletion policies~\cite{googleplay_user_data_policy}. Yet many MAR policies in our dataset omit disclosures aligned with these requirements. App markets could integrate automated auditing tools into pre-release or periodic review workflows to flag incomplete, stale, or internally inconsistent privacy policies. Such tools should not replace legal review, but they can provide scalable screening for missing disclosures and high-risk policy gaps.

\textbf{For MAR app developers.}
Developers should treat privacy policies as maintained artifacts that evolve with app functionality and data practices. Developers should update policies when app functionality changes, periodically review whether policies remain aligned with current legal requirements, and avoid relying solely on generic APPG templates. For MAR apps in particular, privacy policies should explicitly disclose sensor-derived and AR-specific data practices, including spatial, environmental, biometric-adjacent, and retention-related practices where applicable.

\vspace{-0.3cm}
\subsection{Limitations}
Our study focuses on the static documentation of privacy policies. We do not perform code-level static analysis or dynamic program analysis to verify whether runtime app behavior matches stated policy disclosures. This limitation is important because the MAR ecosystem presents methodological hurdles that are distinct from those of general mobile apps.

Verifying MAR app behavior beyond policy text is particularly challenging. First, as highlighted in Section \ref{sec:dataset-overview}, the diversity and rapid evolution of AR SDKs complicate static program analysis. Many MAR apps rely on proprietary, poorly documented, or frequently changing AR libraries, making it difficult to maintain reliable signatures or taint-tracking models for AR-specific data flows. Second, dynamic analysis requires identifying the runtime entry points that activate AR functionality, which may be embedded in app-specific UI flows, camera modes, product viewers, games, or educational interactions. Even after such entry points are identified, triggering the relevant behavior may require specific objects, faces, surfaces, locations, or user interactions. These requirements make semi-automated or fully automated large-scale testing substantially more difficult for MAR apps than for ordinary mobile apps.

\vspace{-0.3cm}
\section{Conclusion}
In recent years, the rapid development of the metaverse and the increasing adoption of MAR have raised significant privacy and security concerns. However, due to the lack of dedicated datasets, the privacy-policy disclosure landscape within the MAR ecosystem remains insufficiently studied. We present the first large-scale empirical audit of MAR privacy policies under enacted U.S.\ state comprehensive privacy laws. Our findings reveal widespread and systematic disclosure deficiencies across multiple dimensions of the ecosystem, showing that MAR privacy policies have not kept pace with either the sensitivity of AR data practices or the growing complexity of U.S.\ state privacy regulation. By releasing our dataset, taxonomy, and auditing pipeline, we aim to support future research on reproducible MAR security and privacy measurement and scalable privacy-policy auditing.

\bibliographystyle{IEEEtran}
\vspace{-0.3cm}
\bibliography{reference}

@article{li2024xrzoo,
  author       = {Shuqing Li and
                  Chenran Zhang and
                  Cuiyun Gao and
                  Michael R. Lyu},
  title        = {{XRZoo}: {A} Large-Scale and Versatile Dataset of Extended Reality {(XR)}
                  Applications},
  journal      = {CoRR},
  volume       = {abs/2412.06759},
  year         = {2024},
  doi          = {10.48550/ARXIV.2412.06759},
  eprinttype   = {arXiv},
  eprint       = {2412.06759},
}

@article{Alghamdi2025xrdroid,
  author       = {Abdulaziz Alghamdi and
                  Ali Al Kinoon and
                  Ahod Alghuried and
                  David Mohaisen},
  title        = {xr-droid: {A} Benchmark Dataset for {AR/VR} and Security Applications},
  journal      = {{IEEE} Trans. Dependable Secur. Comput.},
  volume       = {22},
  number       = {2},
  pages        = {1418--1430},
  year         = {2025},
  doi          = {10.1109/TDSC.2024.3440662},
}

@inproceedings{Zhan2024vpvet,
  author       = {Yuxia Zhan and
                  Yan Meng and
                  Lu Zhou and
                  Yichang Xiong and
                  Xiaokuan Zhang and
                  Lichuan Ma and
                  Guoxing Chen and
                  Qingqi Pei and
                  Haojin Zhu},
  title        = {{VPVet}: Vetting Privacy Policies of Virtual Reality Apps},
  booktitle    = {Proc. {ACM} {SIGSAC} Conf. Comput. Commun. Secur. ({CCS})},
  pages        = {1746--1760},
  month        = oct,
  year         = {2024},
  doi          = {10.1145/3658644.3690321},
}

@inproceedings{Trimananda2022ovrseen,
  author       = {Rahmadi Trimananda and
                  Hieu Le and
                  Hao Cui and
                  Janice Tran Ho and
                  Anastasia Shuba and
                  Athina Markopoulou},
  title        = {{OVRseen}: Auditing Network Traffic and Privacy Policies in {Oculus} {VR}},
  booktitle    = {Proc. {USENIX} Secur. Symp.},
  pages        = {3789--3806},
  month        = aug,
  year         = {2022},
}

@inproceedings{kocmi2025WMT,
    title = "Findings of the {WMT}25 General Machine Translation Shared Task: Time to Stop Evaluating on Easy Test Sets",
    author = "Kocmi, Tom  and
      Artemova, Ekaterina  and
      Avramidis, Eleftherios  and
      Bawden, Rachel  and
      Bojar, Ond{\v{r}}ej  and
      Dranch, Konstantin  and
      Dvorkovich, Anton  and
      Dukanov, Sergey  and
      Fishel, Mark  and
      Freitag, Markus  and
      Gowda, Thamme  and
      Grundkiewicz, Roman  and
      Haddow, Barry  and
      Karpinska, Marzena  and
      Koehn, Philipp  and
      Lakougna, Howard  and
      Lundin, Jessica  and
      Monz, Christof  and
      Murray, Kenton  and
      Nagata, Masaaki  and
      Perrella, Stefano  and
      Proietti, Lorenzo  and
      Popel, Martin  and
      Popovi{\'c}, Maja  and
      Riley, Parker  and
      Shmatova, Mariya  and
      Steingr{\'i}msson, Steinth{\'o}r  and
      Yankovskaya, Lisa  and
      Zouhar, Vil{\'e}m",
    booktitle = "Proc. Conf. Mach. Transl. ({WMT})",
    month = nov,
    year = "2025",
    doi = "10.18653/v1/2025.wmt-1.22",
    pages = "355--413",
}

@inproceedings{Alecci2024androzoo,
  author       = {Marco Alecci and
                  Pedro Jes{\'{u}}s Ruiz Jim{\'{e}}nez and
                  Kevin Allix and
                  Tegawend{\'{e}} F. Bissyand{\'{e}} and
                  Jacques Klein},
  title        = {{AndroZoo}: {A} Retrospective with a Glimpse into the Future},
  booktitle    = {Proc. {IEEE/ACM} Int. Conf. Mining Softw. Repositories ({MSR})},
  pages        = {389--393},
  month        = apr,
  year         = {2024},
  doi          = {10.1145/3643991.3644863},
}

@inproceedings{Farr2025red-ct,
  author       = {David Farr and
                  Nico Manzonelli and
                  Iain Cruickshank and
                  Jevin West},
  title        = {{RED-CT:} {A} Systems Design Methodology for Using {LLM}-labeled Data
                  to Train and Deploy Edge Linguistic Classifiers},
  booktitle    = {Proc. Int. Conf. Comput. Linguistics ({COLING})},
  pages        = {58--67},
  month        = jan,
  year         = {2025},
}

@inproceedings{He2023DeBERTaV3,
  author       = {Pengcheng He and
                  Jianfeng Gao and
                  Weizhu Chen},
  title        = {{DeBERTaV3}: Improving {DeBERTa} using {ELECTRA}-Style Pre-Training with
                  Gradient-Disentangled Embedding Sharing},
  booktitle    = {Proc. Int. Conf. Learn. Representations ({ICLR})},
  month        = may,
  year         = {2023},
}

@inproceedings{Xiang2023policychecker,
  author       = {Anhao Xiang and
                  Weiping Pei and
                  Chuan Yue},
  title        = {{PolicyChecker}: Analyzing the {GDPR} Completeness of Mobile Apps' Privacy
                  Policies},
  booktitle    = {Proc. {ACM} {SIGSAC} Conf. Comput. Commun. Secur. ({CCS})},
  pages        = {3373--3387},
  month        = nov,
  year         = {2023},
  doi          = {10.1145/3576915.3623067},
}

@inproceedings{Xie2025evaluating,
  author       = {Qinge Xie and
                  Karthik Ramakrishnan and
                  Frank Li},
  title        = {Evaluating Privacy Policies under Modern Privacy Laws At Scale: An
                  {LLM}-Based Automated Approach},
  booktitle    = {Proc. {USENIX} Secur. Symp.},
  pages        = {5797--5816},
  month        = aug,
  year         = {2025},
}

@misc{qwen35,
  author = {{Qwen Team}},
  title  = {{Qwen3.5}: Towards Native Multimodal Agents},
  year   = {2026},
  month  = feb,
  url    = {https://qwen.ai/blog?id=qwen3.5},
  note   = {Blog post, accessed Apr.~20, 2026}
}

@misc{artificialanalysis_leaderboard,
  author = {{Artificial Analysis}},
  title  = {{LLM Leaderboard - Comparison of over 100 AI models from OpenAI, Google, DeepSeek \& others}},
  year   = {2026},
  url    = {https://artificialanalysis.ai/leaderboards/models},
  note   = {Accessed: 2026-04-20}
}

@article{Zhang2025qwen3embedding,
  author       = {Yanzhao Zhang and
                  Mingxin Li and
                  Dingkun Long and
                  Xin Zhang and
                  Huan Lin and
                  Baosong Yang and
                  Pengjun Xie and
                  An Yang and
                  Dayiheng Liu and
                  Junyang Lin and
                  Fei Huang and
                  Jingren Zhou},
  title        = {{Qwen3} Embedding: Advancing Text Embedding and Reranking Through Foundation
                  Models},
  journal      = {CoRR},
  volume       = {abs/2506.05176},
  year         = {2025},
  doi          = {10.48550/ARXIV.2506.05176},
  eprinttype   = {arXiv},
  eprint       = {2506.05176},
}

@article{Cao2023MAR_SDK,
  author       = {Jacky Cao and
                  Kit{-}Yung Lam and
                  Lik{-}Hang Lee and
                  Xiaoli Liu and
                  Pan Hui and
                  Xiang Su},
  title        = {Mobile Augmented Reality: User Interfaces, Frameworks, and Intelligence},
  journal      = {{ACM} Comput. Surv.},
  volume       = {55},
  number       = {9},
  pages        = {189:1--189:36},
  year         = {2023},
  doi          = {10.1145/3557999},
}

@article{Javed2024pp_review,
  author       = {Yousra Javed and
                  Ayesha Sajid},
  title        = {A Systematic Review of Privacy Policy Literature},
  journal      = {{ACM} Comput. Surv.},
  volume       = {57},
  number       = {2},
  pages        = {45:1--45:43},
  year         = {2025},
  doi          = {10.1145/3698393},
}

@inproceedings{Herskovitz2020iOS,
  author       = {Jaylin Herskovitz and
                  Jason Wu and
                  Samuel White and
                  Amy Pavel and
                  Gabriel Reyes and
                  Anhong Guo and
                  Jeffrey P. Bigham},
  title        = {Making Mobile Augmented Reality Applications Accessible},
  booktitle    = {Proc. {ACM} {SIGACCESS} Conf. Comput. Accessibility ({ASSETS})},
  pages        = {3:1--3:14},
  month        = oct,
  year         = {2020},
  doi          = {10.1145/3373625.3417006},
}

@inproceedings{YangZ22MAR,
  author       = {Xiaoyi Yang and
                  Xueling Zhang},
  title        = {A Study of User Privacy in {Android} Mobile {AR} Apps},
  booktitle    = {Proc. {IEEE/ACM} Int. Conf. Autom. Softw. Eng. ({ASE})},
  pages        = {226:1--226:5},
  month        = oct,
  year         = {2022},
  doi          = {10.1145/3551349.3560512},
}

@inproceedings{Guo24OculusVR,
  author       = {Hanyang Guo and
                  Hong{-}Ning Dai and
                  Xiapu Luo and
                  Zibin Zheng and
                  Gengyang Xu and
                  Fengliang He},
  title        = {An Empirical Study on {Oculus} Virtual Reality Applications: Security
                  and Privacy Perspectives},
  booktitle    = {Proc. {IEEE/ACM} Int. Conf. Softw. Eng. ({ICSE})},
  pages        = {159:1--159:13},
  month        = apr,
  year         = {2024},
  doi          = {10.1145/3597503.3639082},
}

@inproceedings{Harkous2018polisis,
  author       = {Hamza Harkous and
                  Kassem Fawaz and
                  R{\'{e}}mi Lebret and
                  Florian Schaub and
                  Kang G. Shin and
                  Karl Aberer},
  title        = {{Polisis}: Automated Analysis and Presentation of Privacy Policies Using
                  Deep Learning},
  booktitle    = {Proc. {USENIX} Secur. Symp.},
  pages        = {531--548},
  month        = aug,
  year         = {2018},
}

@inproceedings{Qiu2023Calpric,
  author       = {Wenjun Qiu and
                  David Lie and
                  Lisa M. Austin},
  title        = {{Calpric}: Inclusive and Fine-grain Labeling of Privacy Policies with
                  Crowdsourcing and Active Learning},
  booktitle    = {Proc. {USENIX} Secur. Symp.},
  pages        = {1055--1072},
  month        = aug,
  year         = {2023},
}

@inproceedings{Andow2019policylint,
  author       = {Benjamin Andow and
                  Samin Yaseer Mahmud and
                  Wenyu Wang and
                  Justin Whitaker and
                  William Enck and
                  Bradley Reaves and
                  Kapil Singh and
                  Tao Xie},
  title        = {{PolicyLint}: Investigating Internal Privacy Policy Contradictions on
                  {Google Play}},
  booktitle    = {Proc. {USENIX} Secur. Symp.},
  pages        = {585--602},
  month        = aug,
  year         = {2019},
}

@inproceedings{Andow2020policheck,
  author       = {Benjamin Andow and
                  Samin Yaseer Mahmud and
                  Justin Whitaker and
                  William Enck and
                  Bradley Reaves and
                  Kapil Singh and
                  Serge Egelman},
  title        = {Actions Speak Louder than Words: Entity-Sensitive Privacy Policy and
                  Data Flow Analysis with {PoliCheck}},
  booktitle    = {Proc. {USENIX} Secur. Symp.},
  pages        = {985--1002},
  month        = aug,
  year         = {2020},
}

@inproceedings{Manandhar2022smarthome,
  author       = {Sunil Manandhar and
                  Kaushal Kafle and
                  Benjamin Andow and
                  Kapil Singh and
                  Adwait Nadkarni},
  title        = {Smart Home Privacy Policies Demystified: {A} Study of Availability,
                  Content, and Coverage},
  booktitle    = {Proc. {USENIX} Secur. Symp.},
  pages        = {3521--3538},
  month        = aug,
  year         = {2022},
}

@inproceedings{Adhikari2025policypluse,
  author       = {Andrick Adhikari and
                  Sanchari Das and
                  Rinku Dewri},
  title        = {{PolicyPulse}: Precision Semantic Role Extraction for Enhanced Privacy
                  Policy Comprehension},
  booktitle    = {Proc. Netw. Distrib. Syst. Secur. Symp. ({NDSS})},
  month        = feb,
  year         = {2025},
}

@inproceedings{Cui2023poligraph,
  author       = {Hao Cui and
                  Rahmadi Trimananda and
                  Athina Markopoulou and
                  Scott Jordan},
  title        = {{PoliGraph}: Automated Privacy Policy Analysis using Knowledge Graphs},
  booktitle    = {Proc. {USENIX} Secur. Symp.},
  pages        = {1037--1054},
  month        = aug,
  year         = {2023},
}

@inproceedings{mori2025evaluating,
  author       = {Keika Mori and
                  Daiki Ito and
                  Takumi Fukunaga and
                  Takuya Watanabe and
                  Yuta Takata and
                  Masaki Kamizono and
                  Tatsuya Mori},
  title        = {Evaluating {LLMs} Towards Automated Assessment of Privacy Policy
                  Understandability},
  booktitle    = {Proc. Symp. Usable Secur. Privacy ({USEC})},
  year         = {2025},
  month        = feb,
  doi          = {10.14722/usec.2025.23009}
}

@article{Lehman2022riksofMAR,
  author       = {Sarah M. Lehman and
                  Abrar S. Alrumayh and
                  Kunal Kolhe and
                  Haibin Ling and
                  Chiu C. Tan},
  title        = {Hidden in Plain Sight: Exploring Privacy Risks of Mobile Augmented
                  Reality Applications},
  journal      = {{ACM} Trans. Priv. Secur.},
  volume       = {25},
  number       = {4},
  pages        = {26:1--26:35},
  year         = {2022},
  doi          = {10.1145/3524020},
}

@inproceedings{Pan2024APPGs,
  author       = {Shidong Pan and
                  Dawen Zhang and
                  Mark Staples and
                  Zhenchang Xing and
                  Jieshan Chen and
                  Xiwei Xu and
                  Thong Hoang},
  title        = {Is It a Trap? {A} Large-scale Empirical Study And Comprehensive Assessment
                  of Online Automated Privacy Policy Generators for Mobile Apps},
  booktitle    = {Proc. {USENIX} Secur. Symp.},
  month        = aug,
  year         = {2024},
}

@misc{googleplay_user_data_policy,
  author       = {{Google}},
  year = {2026},
  title        = {{User Data Policy}},
  howpublished = {\url{https://support.google.com/googleplay/android-developer/answer/10144311}},
  note         = {{Google Play} Developer Policy Center. Accessed: 2026-04-25}
}

@misc{Snapchat,
  title        = {{Boone et al. v. Snap Inc.: Class Action Complaint}},
  author       = {{Boone et al.}},
  year         = {2022},
  month        = aug,
  day          = {4},
  howpublished = {Circuit Court of the Eighteenth Judicial Circuit, DuPage County, Illinois, Case No. 2022-LA-708},
  url          = {https://angeion-public.s3.amazonaws.com/www.SnapIllinoisBIPASettlement.com/docs/Snap-Class%20Action%20Complaint.pdf},
  urldate      = {2026-04-27}
}

@misc{Tiktok,
  title        = {{In re TikTok, Inc., Consumer Privacy Litigation: Consolidated Amended Class Action Complaint}},
  author       = {{Plaintiffs in In re TikTok, Inc., Consumer Privacy Litigation}},
  year         = {2020},
  month        = dec,
  day          = {18},
  howpublished = {United States District Court for the Northern District of Illinois, Eastern Division, MDL No. 2948, Master Docket No. 20-cv-4699, Document 114},
  url          = {https://cdn.arstechnica.net/wp-content/uploads/2021/02/amended-complaint-tiktok-consumer-privacy-litigation.pdf},
  urldate      = {2026-04-27}
}

@misc{CharlotteTilbury,
  title        = {{Halim v. Charlotte Tilbury Beauty Inc. et al.: Class Action Complaint}},
  author       = {{Halim, Olena}},
  year         = {2022},
  month        = dec,
  day          = {7},
  howpublished = {Circuit Court of Cook County, Illinois, Case No. 2022-CH-11832},
  url          = {https://www.classaction.org/media/halim-v-charlotte-tilbury-beauty-inc-et-al.pdf},
  urldate      = {2026-04-27}
}

@misc{LouisVuitton,
  title        = {{Theriot v. Louis Vuitton North America, Inc.: Class Action Complaint}},
  author       = {{Theriot, Paula}},
  year         = {2022},
  month        = apr,
  day          = {8},
  howpublished = {United States District Court for the Southern District of New York, Case No. 1:22-cv-02944},
  url          = {https://www.classaction.org/media/theriot-v-louis-vuitton-north-america-inc.pdf},
  urldate      = {2026-04-27}
}

@misc{iapp_state_privacy_tracker_2026,
  title        = {{US State Privacy Legislation Tracker}},
  author       = {{International Association of Privacy Professionals}},
  year         = {2026},
  howpublished = {IAPP Resource},
  url          = {https://iapp.org/resources/article/us-state-privacy-legislation-tracker},
  urldate      = {2026-04-27},
  note         = {Dynamic tracker of comprehensive U.S. state privacy bills and laws; statistics counted by the authors as of January 1, 2026}
}

@misc{statista_mobile_ar_users,
  title        = {Mobile augmented reality ({AR}) users worldwide 2023--2028},
  author       = {Taylor, Petroc},
  year         = {2025},
  month        = nov,
  day          = {27},
  howpublished = {\url{https://www.statista.com/statistics/1098630/global-mobile-augmented-reality-ar-users/}},
  note         = {Accessed: 2026-04-27}
}

@misc{googletrans,
  author = {ssut},
  title = {{googletrans}},
  year = {2025},
  url = {https://github.com/ssut/py-googletrans},
  note = {Accessed: 2026-04-28}
}

@misc{googleplayscraper,
  author = {facundoolano},
  title = {{google-play-scraper}: {Node.js} scraper to get data from {Google Play}},
  year = {2025},
  howpublished = {\url{https://github.com/facundoolano/google-play-scraper}},
  note = {{MIT} License; accessed 2026-04-28}
}

@misc{playwright,
  author = {{Microsoft}},
  title = {Playwright},
  year = {2025},
  howpublished = {\url{https://github.com/microsoft/playwright}},
  note = {Accessed: 2026-04-28}
}

@misc{singlefile,
  author = {Lormeau, Gildas},
  title = {{SingleFile}: Save a complete web page as a single {HTML} file},
  year = {2025},
  howpublished = {\url{https://github.com/gildas-lormeau/SingleFile}},
  note = {Accessed: 2026-04-28}
}

@misc{rapidfuzz,
  author = {maxbachmann},
  title = {{RapidFuzz}},
  year = {2025},
  howpublished = {\url{https://github.com/maxbachmann/rapidfuzz}},
  note = {Accessed: 2026-04-28}
}

@misc{html2text,
  author = {Alir3z4},
  title = {{html2text}},
  year = {2025},
  howpublished = {\url{https://github.com/Alir3z4/html2text}},
  note = {Accessed: 2026-04-28}
}

@misc{trafilatura,
  author = {adbar},
  title = {{trafilatura}},
  year = {2025},
  howpublished = {\url{https://github.com/adbar/trafilatura}},
  note = {Accessed: 2026-04-28}
}

@misc{markitdown,
  author = {{Microsoft}},
  title = {{MarkItDown}},
  year = {2025},
  howpublished = {\url{https://github.com/microsoft/markitdown}},
  note = {Accessed: 2026-04-28}
}

@misc{readabilitylxml,
  author = {martinblech},
  title = {{readability-lxml}},
  year = {2025},
  howpublished = {\url{https://pypi.org/project/readability-lxml/}},
  note = {Accessed: 2026-04-28}
}

@inproceedings{Hu2022LoRA,
  author       = {Edward J. Hu and
                  Yelong Shen and
                  Phillip Wallis and
                  Zeyuan Allen{-}Zhu and
                  Yuanzhi Li and
                  Shean Wang and
                  Lu Wang and
                  Weizhu Chen},
  title        = {{LoRA}: Low-Rank Adaptation of Large Language Models},
  booktitle    = {Proc. Int. Conf. Learn. Representations ({ICLR})},
  month        = apr,
  year         = {2022},
}

@misc{apkeep,
  author       = {{Electronic Frontier Foundation}},
  title        = {{apkeep}: A command-line tool for downloading APK files from various sources},
  year         = {2025},
  howpublished = {\url{https://github.com/EFForg/apkeep}},
  note         = {Accessed: 2026-04-28}
}

@misc{sensortower,
  author       = {{Sensor Tower}},
  title        = {{Sensor Tower}: Digital Intelligence and App Data Analysis},
  howpublished = {\url{https://sensortower.com/}},
  note         = {Accessed: 2026-04-28},
  year         = {2026}
}

@inproceedings{Schimanski24llm_specialists,
  author       = {Tobias Schimanski and
                  Jingwei Ni and
                  Mathias Kraus and
                  Elliott Ash and
                  Markus Leippold},
  title        = {Towards Faithful and Robust {LLM} Specialists for Evidence-Based Question-Answering},
  booktitle    = {Proc. Annu. Meeting Assoc. Comput. Linguistics ({ACL})},
  pages        = {1913--1931},
  month        = aug,
  year         = {2024},
  doi          = {10.18653/V1/2024.ACL-LONG.105},
}

@article{ZhangZ25llm_hallucinations_in_code,
  author       = {Ziyao Zhang and
                  Chong Wang and
                  Yanlin Wang and
                  Ensheng Shi and
                  Yuchi Ma and
                  Wanjun Zhong and
                  Jiachi Chen and
                  Mingzhi Mao and
                  Zibin Zheng},
  title        = {{LLM} Hallucinations in Practical Code Generation: Phenomena, Mechanism,
                  and Mitigation},
  journal      = {Proc. {ACM} Softw. Eng.},
  volume       = {2},
  number       = {{ISSTA}},
  pages        = {481--503},
  year         = {2025},
  doi          = {10.1145/3728894},
}

@inproceedings{libscout,
  author       = {Michael Backes and
                  Sven Bugiel and
                  Erik Derr},
  title        = {Reliable Third-Party Library Detection in {Android} and its Security
                  Applications},
  booktitle    = {Proc. {ACM} {SIGSAC} Conf. Comput. Commun. Secur. ({CCS})},
  pages        = {356--367},
  month        = oct,
  year         = {2016},
  doi          = {10.1145/2976749.2978333},
}

@inproceedings{zhong2013googleplay,
  author       = {Nan Zhong and
                  Florian Michahelles},
  title        = {{Google Play} is not a long tail market: an empirical analysis of app
                  adoption on the {Google Play} app market},
  booktitle    = {Proc. {ACM} Symp. Appl. Comput. ({SAC})},
  pages        = {499--504},
  month        = mar,
  year         = {2013},
  doi          = {10.1145/2480362.2480460},
}

@article{HarborthP2021MAR,
  author       = {David Harborth and
                  Sebastian Pape},
  title        = {Investigating privacy concerns related to mobile augmented reality
                  Apps - {A} vignette based online experiment},
  journal      = {Comput. Hum. Behav.},
  volume       = {122},
  pages        = {106833},
  year         = {2021},
  doi          = {10.1016/J.CHB.2021.106833},
}

@inproceedings{Tang2020ios,
  author       = {Zhushou Tang and
                  Ke Tang and
                  Minhui Xue and
                  Yuan Tian and
                  Sen Chen and
                  Muhammad Ikram and
                  Tielei Wang and
                  Haojin Zhu},
  title        = {{iOS}, Your {OS}, Everybody's {OS}: Vetting and Analyzing Network Services
                  of {iOS} Applications},
  booktitle    = {Proc. {USENIX} Secur. Symp.},
  pages        = {2415--2432},
  month        = aug,
  year         = {2020},
}

@misc{StatCounter2026MobileOS,
  author       = {{StatCounter}},
  title        = {{Mobile Operating System Market Share Worldwide}},
  howpublished = {\url{https://gs.statcounter.com/os-market-share/mobile/worldwide}},
  note         = {Accessed: 2026-04-28},
  year         = {2026}
}

\vfill

\end{document}